# Generation of composite vortex beams by independent Spatial Light Modulator pixel addressing

September 2019

**Authors:**

Mateusz Szatkowski*[a], Jan Masajada[a], Ireneusz Augustyniak[b], Klaudia Nowacka[a].

**Addresses:**

[a] Department of Optics and Photonics, Wroclaw University of Science and Technology, Wyb. Wyspianskiego 27, 50-370 Wroclaw, Poland

[b] Department of Applied Mathematics, Wrocław University of Science and Technology, Wybrzeże Wyspiańskiego 27, 50-370 Wrocław, Poland

*mateusz.szatkowski@pwr.edu.pl

**ABSTRACT**

The composite optical beams being a result of superposition, are a promising way to study the orbital angular momentum and its effects. Their wide range of applications makes them attractive and easily available due to the growing interest in the Spatial Light Modulators (SLM). In this paper, we present a simple method for generating composite vortex patterns with high symmetry. Our method is simple, flexible and gives perfectly aligned beams, insensitive to mechanical vibrations. This method is based on the ability to split SLM cells between phase patterns that are to be superposed. This approach allows control of the intensity relation between those structures, enables their rotation and is capable to superpose more than two such structures. In this paper, we examine its ability to produce superposition of two optical vortices by presenting both theoretical and experimental results.

1. **INTRODUCTION**

The spatial light modulators (SLMs) [1] are used for wavefront geometry shaping. This results in a growing number of SLM applications, for instance: laser beam shaping [2], adaptive optics [3–5], image processing [6], optical trapping [7–9], dynamic holography and image projection [10,11]. Together with a wide range of applications, some serious drawbacks of the SLM are present, concerning their optical quality, stability, finite range of the phase modulation they are capable of, relatively low speed and memory effects [1]. Fortunately, some of those can be corrected sufficiently for many applications [12–15]. Having well-corrected SLM it is possible to generate beams belonging to the well-known families such as Gaussian vortex beams [16,17], Bessel beams [18], Airy beams [19]. Vector beam generation is also achievable [20,21]. SLM can modify the standard forms of the given beam to improve its usefulness. The improved vortex beams - so-called perfect vortex beams [22,23] are a good example. There are also some papers devoted to more complicated patterns, which can be produced with the SLM [24–26]. These combinations of various vortex beams are usually produced just by the superposition of two separated beams. Several papers are referring to this method [27–33]. Some of them require very stable conditions, i.e. both beams should overlap perfectly while propagating along the z-axis. A good example is a theoretical paper [27] where the angular momentum of the beam carrying an optical vortex is analyzed in analogy with rigid body mechanics. The vortex beam was obtained by interference of Gaussian and Laguerre-Gaussian mode. The experiment, which could reproduce the theoretical results, requires prefect and stable beam alignments.

In this paper, we study a simple method for generating many composite vortex patterns with high symmetry. Our method is simple and flexible. Both beams are generated with single SLM, so in the result, they propagate along the same axis. Thus, the beam alignment does not change under mechanical vibrations. The generated patterns can be easily rotated, and the intensity of various details can be modified. We split the SLM cells randomly between the two-phase patterns (Figure 1) which are to be mixed (by interference). A similar method of random (or deterministic) pixel division has been already used in a wide range of applications, for example, holographic data encoding [34] or for producing controllable speckle fields [35]. In the field of structured light, random pixel division is generally used to manipulate or redirect the unmodulated part of the beam, improving the perfect vortex beam is a good example [36]. However, the applications mentioned above, do not collide with our work. We do not improve the quality of the beam generated by SLM but interfere two of them to produce a new shape and investigate the general properties of the used method.

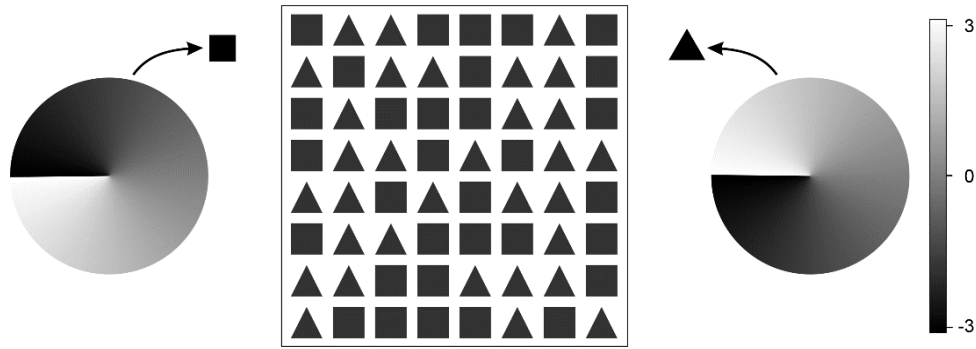

**Figure 1.** Filling scheme for the composite vortex pattern. The SLM pixels are split into two groups marked by triangles and squares. The two vortex phase patterns with opposite topological charges are written onto the SLM in such a way that one vortex pattern occupies pixels marked by triangles and the other vortex patterns occupy pixels marked by squares.

The other approach is dividing the SLM surface into two or more compact parts (for example into two equal parts or four equal quadrants) [6,37–39]. Considering the result (i.e. obtained images) this method is not always equivalent to ours. However, in many cases, our method can do the same more efficiently and elegantly

## 2. THEORY

We divide the SLM's pixels into two interlaced subsets so that the neighboring SLM cells are used for displaying various phase patterns. Both phase patterns are used for shaping the reflected beam. These two beams interfere at the image (camera) plane. The other way is a random pixel splitting (Figure 1). We have tested both methods and the differences between the recorded interference images are negligible. In the following sections, we will present such images obtained with a random splitting method, due to its higher flexibility. In the case of random pixels distribution, the number of pixels in subsequent parts does not have to be equal. We can easily change the ratio of pixels assigned to the two-phase patterns. For example, one pattern may have 70% of the total number of pixels while the second one 30%. As we will show, this is important for equaling the light distribution between the bright and dark parts of the image.

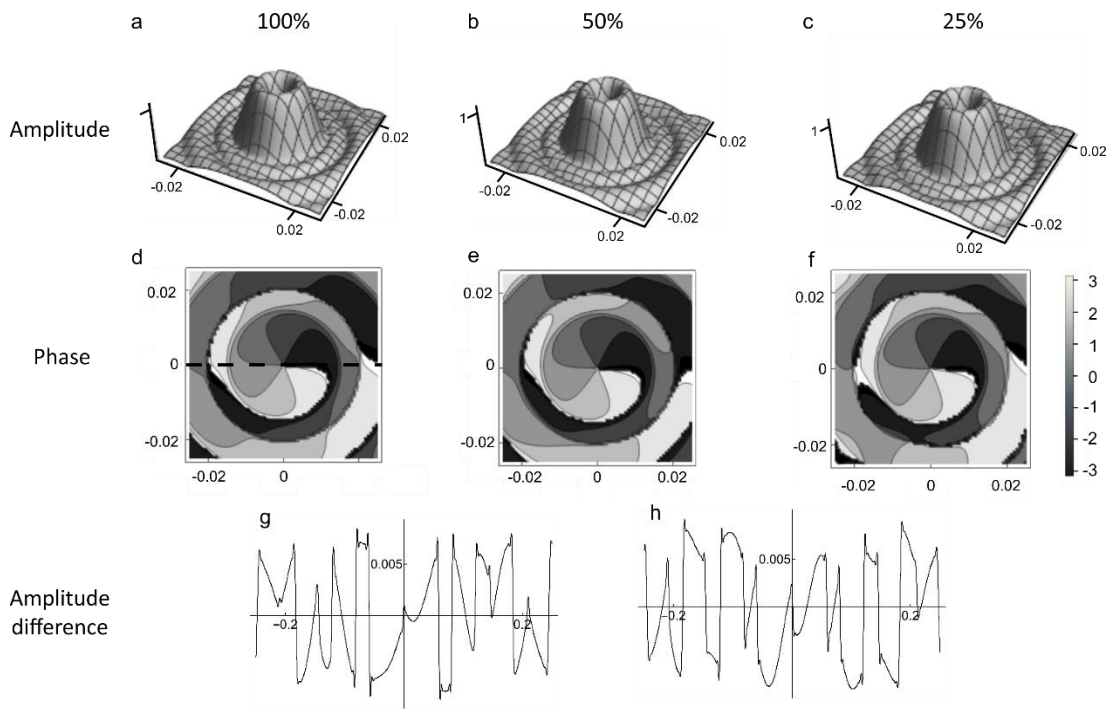

**Figure 1.** Numerical analysis of the images generated by fraction of the SLM pixels: a) – c) show the amplitude and e) – f) phase distribution in the case of: a), d) 100% b), e) 50% and c), f) 25% pixels used for displaying the phase vortex pattern. g) shows the amplitude difference between b) and a) and h) the same, but between c) and a). These differences were calculated along the thick-dashed line shown d).

By taking every second cell of the SLM we decrease their number and enlarge the distance between the neighboring cells belonging to the same group. However, the total number of cells of the currently available SLM is large enough for getting images of good quality, even if the fraction of them is used. Figure 2a-c shows the image amplitude distribution calculated numerically in the case of three pixels group used for displaying the single vortex pattern, (a - all pixels, b - half of all pixels, c - quarter of all pixels). In our calculations, we assume that the SLM is illuminated by the plane wave. After passing through the SLM the light goes through the lens with focal length f=100mm. The image plane is located at the focal plane of the lens.

The difference between the image, produced by all SLM cells and a part of them is small. The phase patterns Figure 2d-f differ in tiny details. The differences between amplitude distribution are also negligible (Figure 2g-h) and in most practical cases can be neglected. This means that we can compute our images as a result of the interference of two beams generated independently by the two groups of the SLM cells. Having this in mind we can start to study the resulting images by adding two complex amplitudes describing two optical vortices of opposite topological charge. The two different images originating from the two parts of the SLM interfere at the image plane, which can be calculated just by adding their complex amplitudes. We add two vortex beams having Gaussian amplitude envelope and opposite topological charges. In the polar coordinates $(\rho, \varphi)$ the sum is given by:.

$$u(\rho,\varphi) = \Omega_1 e^{-\frac{\rho^2}{w_{z1}^2}+ik\frac{\rho^2}{2z_{R1}}} e^{-im_1\varphi} + \Omega_2 e^{-\frac{\rho^2}{w_{z2}^2}+ik\frac{\rho^2}{2z_{R2}}} e^{-im_2\varphi} \qquad (1)$$

$$z_{R1} = \frac{kw_{01}}{2}; \quad z_{R2} = \frac{kw_{02}}{2}; \qquad (1a)$$

$$w_{z1}^2 = w_{01}^2\sqrt{1+\frac{z^2}{z_{R1}^2}}; \quad w_{z2}^2 = w_{02}^2\sqrt{1+\frac{z^2}{z_{R2}^2}} \qquad (1b)$$

Where: $\Omega_1$, $\Omega_2$, are complex constant neglected in further calculations, $k$ is a wavenumber, $w_{01}$, $w_{02}$ are Gaussian beam waists, $z_{R1}$, $z_{R2}$ are Rayleigh ranges and $m_1$, $m_2$ are integer numbers defining optical vortex topological charge, $z$ is a distance between waist plane and the plane of observation, indexes 1 and 2 corresponds to the first and second vortex pattern, respectively. The image calculated in this way is shown in Figure 2a-b.

In the second approach we assume that the Gaussian beam $u_G(\rho,\varphi)$ illuminates the phase vortex pattern (VP) and focusing lens:

$$u_G(\rho,\varphi) = \Omega e^{-\frac{\rho^2}{w_z^2}+ik\frac{\rho^2}{2z_R}} \qquad (2)$$

The phase function of the beam with optical vortex of charge $\pm m$:

$$t_V(\varphi) = e^{\pm im\varphi} \qquad (3)$$

The lens transmittance function is given by:

$$t_L(\rho) = e^{-i\frac{k}{2f}\rho^2} \qquad (4)$$

Where: $f$ is lens focal length.

The SLM cells are randomly distributed between two vortex patterns. Then we calculate numerically the image at the lens focal plane. This image is shown in Figure 2c-d. The single vortex beam generated in this way (under Gaussian beam illumination) is described by the difference of two modified Bessel functions [40] instead of formula (1). As a result, the images are shown in Figure 2a-b and Figure 2c-d differ in size. The two vortex beams (positive and negative) added in the formula (1) slightly diverge. After passing through the lens, the Gaussian beam (2) converges, which is responsible for the spot size difference (Figure 2a and c). The image in Figure 2c-d is calculated at the focal plane of the lens where the beam wavefront is almost flat. On the other hand, the beam (1) has larger wavefront curvature. This difference in a wavefront curvature is responsible for the phase distribution differences (Figure 2b and d). Although we add the two vortices of opposite topological charge, both phase images contain a larger vortex number Figure 2b and d). These extra vortices are not surprising as they are present in most of the interference patterns created with the vortex beams [41].

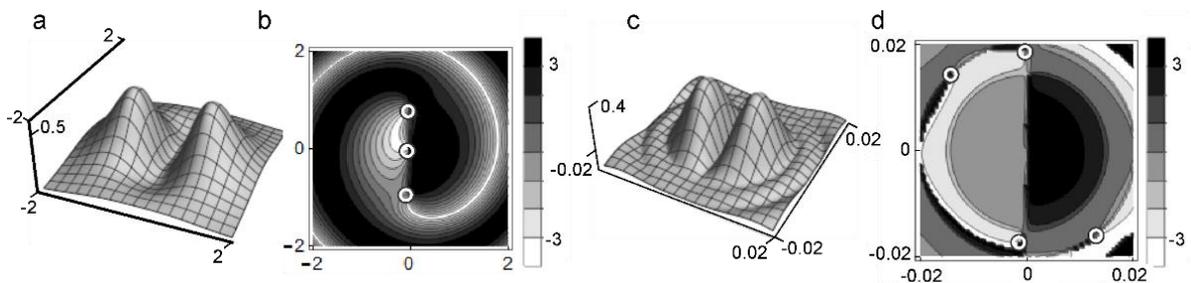

**Figure 2**. Comparison between basic analytical and numerical results. In both cases two vortices of opposite topological charge $m = \pm 1$ are added with the pixel ratio 1:1; a-b, amplitude and phase distribution calculated with the formula (1). The white circles in the image; show the position of three generated vortices c-d, amplitude and phase distribution calculated numerically in the system shown in Figure 1. The white circles show the position of four identified vortices.

The vortex distribution is very sensitive to phase variation introduce into the interfering beams. However, the amplitude distribution given by the formula (1) and obtained after propagation through the VP and lens has the same character. Since we are focused on the general features of the combined vortex images, we will not discuss the details of the phase patterns in this paper. All our conclusions are limited to the interference images as recorded by a camera that cancels the information on phase distribution at the image plane. Below we give a brief discussion of these general features.

The images shown in Figure 2a-c contain two intensity maxima and have broken circular symmetry, contrary to the donut-like shape of the amplitude distribution of optical vortex. This results from the broken symmetry of the total phase pattern displayed on the SLM which is illustrated in Figure 3.

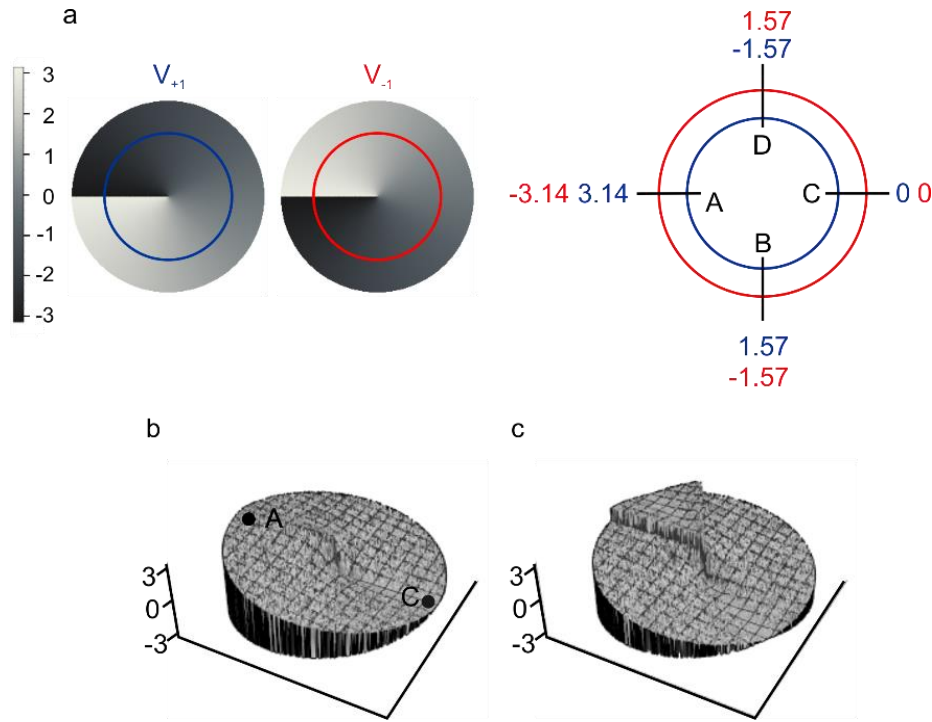

Figure 3. The phase pattern of the composite vortex $V_{+1}$ ($m = +1$) - vortex $V_{-1}$ ($m = -1$) structure. a) the phase pattern of two composite vortices, together with two circles representing phase values of the vortex beam $V_{+1}$ (blue) $V_{-1}$ (red). Some values are explicitly marked by letters. At point C both patterns have phase value 0, at point B the phase value is -1.57 and + 1.57, and at point D +1.57 and -1.57, respectively. At point A the phase value is -π and +π, respectively; b) the 3D image of the phase distribution of composite vortex structure. It consists of needles randomly directed up or down. At point C the needles have value zero and at point A $+\pi$ or $-\pi$. The symmetry of this pattern is lower than the symmetry of the pure vortex state; c) the same but with an additional phase shift by $\pi/3$ of the second ($V_{-1}$) vortex pattern. Thus, the phase of the second vortex is rotated by the $\pi/3$ angle, which results in an additional bump at the total phase pattern.

The interesting feature is a relation between the introduced phase shift to the one vortex pattern (rotation of one vortex pattern) and overall image rotation. When the phase shift by angle $\alpha$ is added, the whole image rotates by angle $\alpha/2$. The phase shift was added to each pixel, representing a particular vortex pattern (displayed by the SLM). Figure 4 shows a series of rotated images and Figure 5 gives the explanation of this effect.

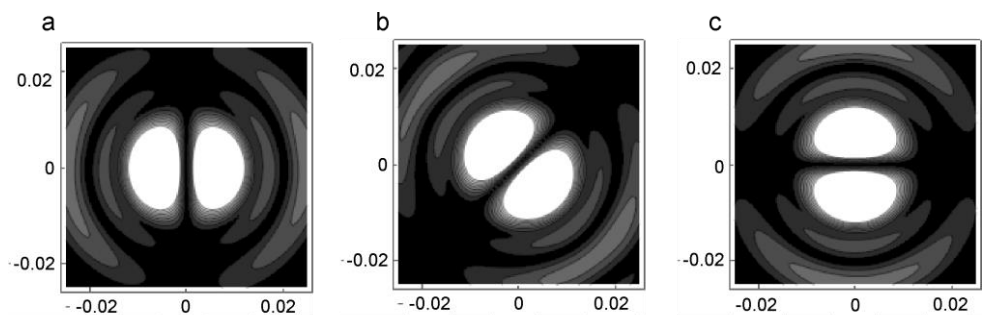

Figure 4. The contour plot of amplitude distribution of the two vortices superposition having topological charge +1 and -1 respectively: a) the vortex phases relation is as in Figure 3a; b) the phase pattern of vortex beam +1 was rotated by $\alpha = \pi/2$ and the image was rotated by $\pi/4$; c) the phase pattern of vortex beam +1 was rotated by $\alpha = \pi$ and the image was rotated by $\pi/2$.

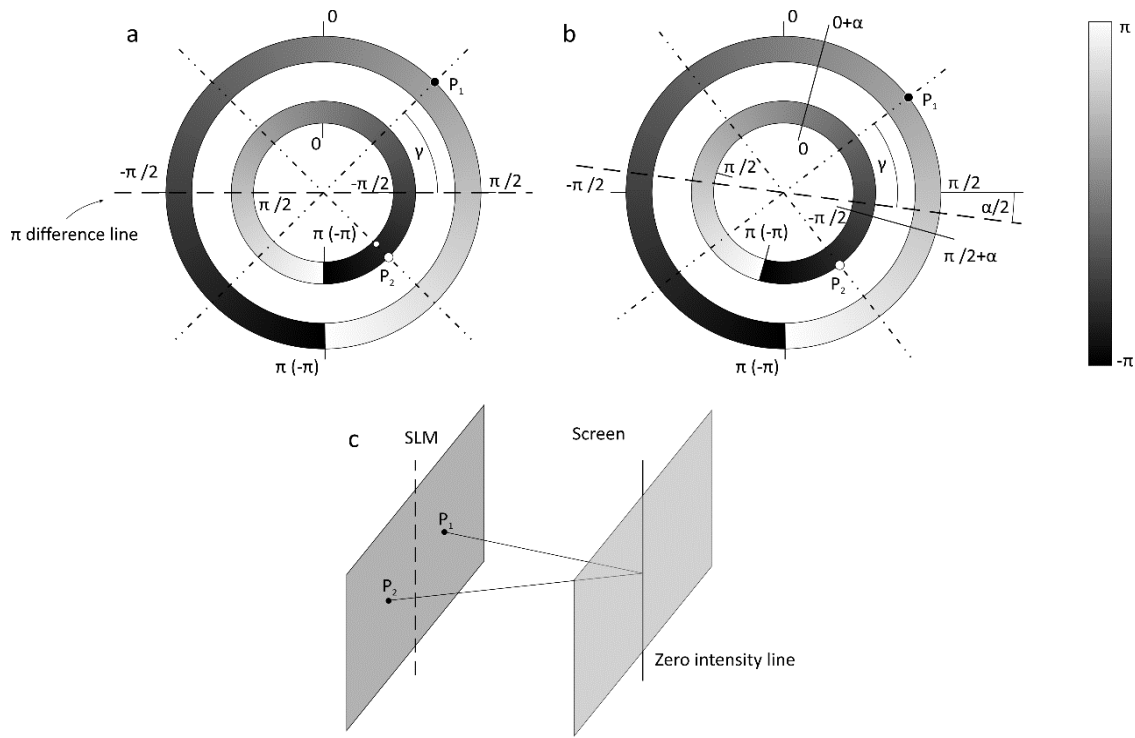

**Figure 5.** The phase relations between two vortex patterns of opposite sign. The two circles in (a) and (b) represent the phase distribution of the vortex pattern A (the outer circle) and B (the inner circle). Both patterns differ in their topological charge sign; a) both patterns have zero phase value in the same direction. In case of vortex A, the phase circulates clockwise and in case of vortex B it circulates counterclockwise. The thick dashed line shows $\pi$- difference line, i.e. the line along which the phase of vortex A and B differs by $\pi$. Also, the phase at points P1 and P2 belonging to the vortex A and B respectively, differs by $\pi$. Due to opposite vortex charges the points P1 and P2 must lay on the lines (dash-dotted lines) rotated by the same angle $\gamma$, but in opposite directions; b) the vortex B was rotated by the angle $\alpha$. Now the $\pi$- difference line has rotated by angle $\alpha/2$, as well as the two dash-dotted lines with points $P_1$ and $P_2$ so that the phase at points $P_1$ and $P_2$ still differ by $\pi$. Each point $P_1$ at the vortex pattern A will have the corresponding point $P_2$ at the vortex pattern B, located symmetrically against the $\pi$-difference line. On this line, both such points collide and become the same one. c) Since the pairs of points $P_1$ and $P_2$ are shifted in phase by $\pi$, there is a line at the screen, positioned symmetrically against these points, along which their contribution cancels. We also assume that the amplitude and phase of the illuminating beam reveal circular symmetry (as Gaussian beam for example), so each pair of points $P_1$ and $P_2$ has the same initial amplitude and phase. While rotating the vortex pattern B by phase angle $\alpha$, the $\pi$-difference line rotates by angle $\alpha/2$ and as a result the whole zero intensity line and the entire image rotates also by $\alpha/2$, as it is visualized in Figure 4.

However, there is still a question of what happens if the vortices have higher, but still opposite topological charges. It is well known that in case of charge $m$ (where $m$ is an integer number) the phase circulates $\pm m2\pi$ times when traveling around the singular point Figure 6a-b. Superposition of two optical vortices with charge $m$ and $-n$ should produce $|m| + |n|$ fold pattern. Figure 6c presents phase change along the circle in Cartesian coordinates. This figure can be interpreted in a dynamic manner. The red lines show the trajectory of the point which starts at the height $2\pi$ and travels down. Whenever it reaches the bottom line at level 0 it jumps immediately back to the top and starts again. Here we have plotted four such passages in time T (it corresponds to the vortex charge $m = -4$). Now we shift the entire red line (i.e. all segments) by $\pi$, but modulo $2\pi$ (violet line), and again we will have four full passages in time T. The black line represents the point that travels by $2\pi$ (in time T) once, but now from the bottom to the top, which corresponds to optical vortex with charge $+1$. Obviously, the black line must cross the violet one five times. We may think that one runner runs down the hill four times faster than the second one walks up. After getting down the first runner (say A) is moved immediately to the top and starts again. The second one (say B) reaches the top at time T. Runner and walker will meet five times at the time T. We must remember that the top and bottom points are the same points on the circle. Otherwise the number of intersection points could be smaller by one. In the next example walker is two times faster (the blue line), so it reaches the top twice in time T. Adding one more run means one more crossing point and now both A and B meet six times at the time T. When the B runs three times from the bottom to top in time T (green), it adds one more segment of $2\pi$ length and in result we have one more crossing point. We can conclude that when adding two vortices of charge $+m$ and $-n$, we will have $|m| + |n|$ crossing points, which will result in $|m| + |n|$ zero lines at the image plane. It is worth noticing here that the violet line was shifted by $\pi$, so the crossing points represent the points where phases of both vortex patterns differ by $\pi$.

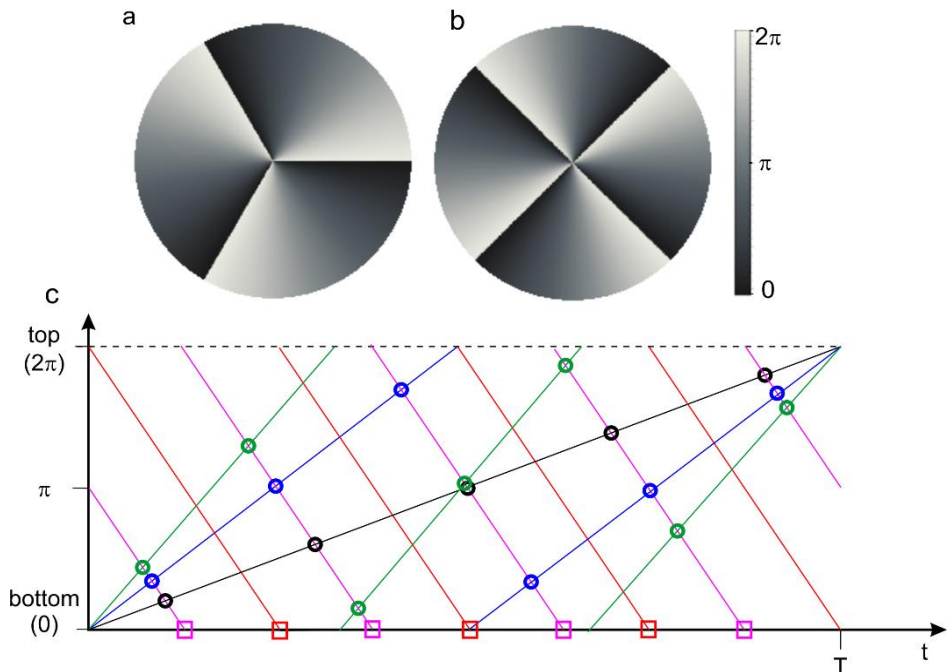

**Figure 6.** Higher-order vortices geometry; a) the phase distribution of the vortex with $m = 3$. The phase circulates clockwise by $6\pi$ around the central point; b) the phase distribution of the vortex with $m = -4$. The phase circulates anticlockwise by $8\pi$ around the central point; c) the red line shows the point traveling by distance $8\pi$ from top to bottom in time T. The violet line shows the same point, but its trajectory is shifted by $\pi$. The red and violet rectangles show the jump points from the bottom to the top. The black line shows the trajectory of a point traveling from the bottom to the top in time T by distance $2\pi$. Black circles show five intersection points between black and violet lines. The blue line shows the point traveling from the bottom to the top by distance $4\pi$. Blue circles show six intersection points between blue and violet lines. The green line shows the point traveling from the bottom to the top by distance $6\pi$. Green circles show seven intersection points between green and violet lines.

Figure 7 shows various combinations of images with seven-fold symmetry. The pattern was obtained by combining vortices of topological charge $m$ and $-n$, which satisfies the condition $|m| + |n| = 7$.

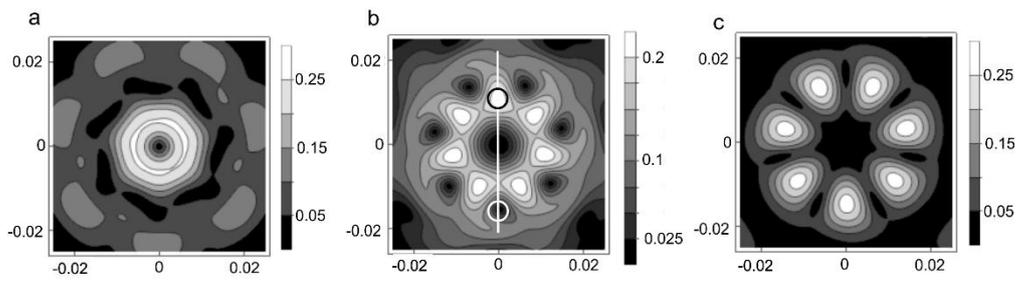

**Figure 7.** Superposition of various vortices, where their topological charges sum $|m|+|n|$ up to seven; a) m=6, n=-1, b) m=5, n=-2; c) m=4, n=-3. The seven-fold geometry is well visible.

Figure 7 requires some comments. We can see that there are no dark lines as it was in the case of combination of vortices with m=1 and m=-1 (Figure 4). There are two reasons for that. The $\pi$- difference line in Figure 5 consists of two parts that start at the vortex pattern center. Crossing the center of the vortex pattern, the phase value jumps from $\pi/2$ and $-\pi/2$ for vortex A and in the opposite direction for vortex B. In fact, we have two $\pi$-difference lines which are responsible for two-fold symmetry of the image. When the $|m| + |n|$ sum is odd, these $\pi$- difference lines go from the center at different directions, so there are no line pairs positioned at opposite directions. Thus, moving along a $\pi$- difference line and crossing the vortex center we find a line along which the phase difference is not equal to $\pi$. In result at the image plane we have bright areas at one side (Figure 7b upper part – in the black circle) and dark areas (but not lines) at the other side (Figure 7b bottom part – white circle), but the $|m| + |n|$ fold symmetry of the entire pattern is preserved. The second reason is that the ring radius of amplitude distribution depends on the topological charge of the vortex. Thus, the contributions from two points being in antiphase do not sum to zero along the symmetry line, when observed at screen, which is illustrated in Figure 8.

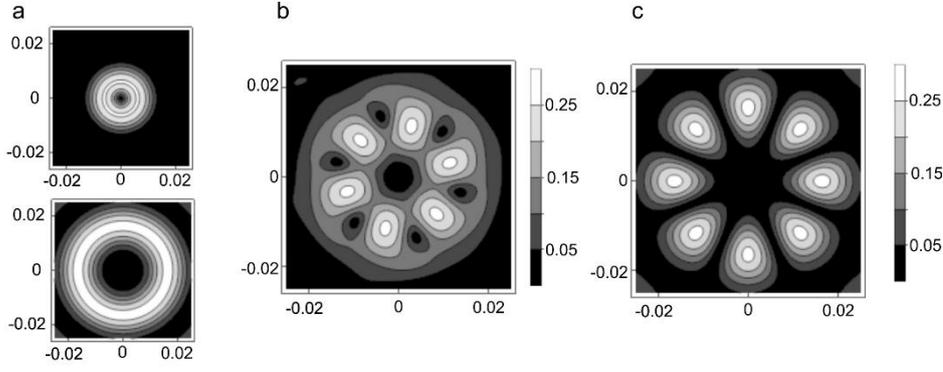

**Figure 8.** The effects of vortex size and parity; a) the amplitude distribution of two vortices (at image plane) top $m = \pm 1$, bottom $m = \pm 4$. The higher vortex charge corresponds to larger amplitude ring diameter; b) the amplitude distribution of the vortex combination $m = 4, n = -2$. Although the sum $|n| + |m|$ is even and the low-intensity areas lie at opposite sides, there are areas and not lines. This is due to various amplitude distribution for vortices having different values of topological charge; c) the amplitude distribution of the vortex combination $m = 4, n = -4$. When both conditions hold, i.e. the sum $|n| + |m|$ is even and $n = -m$ (so the amplitude distribution is the same), there are zero intensity lines going through the center of the image.

There is still the following problem. Assume that we superpose two vortices of topological charge $m$ and $-n$. Next, one of them is rotated by angle $\alpha$, which means that an angle $\alpha$ is added to its phase distribution. By what angle does the whole pattern rotates? The answer in case of $m = 1$ and $n = -1$ was discussed previously (Figure 5). The more general case is presented in topological charge $-n$, and the outer circle is just an angle scale from $-\pi$ to $\pi$ (scaled circle).

Figure 9. The inner circle shows the phase on vortex with topological charge m, the middle shows the phase on vortex with

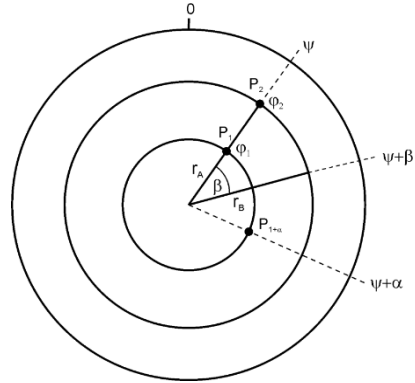

**Figure 9.** Adding two vortex patterns with opposite sign of the topological charge. The inner circle shows the phase values of the vortex pattern A, the middle circles corresponds to the vortex pattern B, and the outer circle is a standard angle scale. The radius $r_A$ is a $\pi$ - difference line. Point $P_1$ belongs to the vortex A and has a phase angle $\varphi_1$ and the point $P_2$ belongs to the vortex B and has the phase angle $\varphi_2$. The angle coordinate of the radius $r_A$ and points $P_1$ and $P_2$ is $\psi$. Adding the phase $\alpha$ means that the whole pattern rotates by $\alpha$ and in result the new coordinate of the point P1 is $\psi + \alpha$. The radius $r_B$ shows a new position of $\pi$ – difference line, which is rotated by the angle $\beta$.

To change the phase of given vortex pattern by the angle $\alpha$ means to add this value to all vortex points. As a result, the whole pattern rotates by angle $\alpha$, measured at the scaled circle. Let the phase difference at points lying along the radius $r_A$ (angle coordinate $\varphi$) of the first and second vortex be $\pi$ (it is a $\pi$ - difference line), then we have:

$$m\psi - n\psi = \varphi_1 - \varphi_2 = \pi \qquad (5)$$

After angle $\alpha$ addition to the first vortex, the phase value along the radius $r_A$ will be changed to the new value $\varphi_1 + \alpha$, and the radius where the condition (5) holds will be moved to the new position $r_B$. The angle between radius $r_A$ and $r_B$ is $\beta$. The phase of the first vortex, along the radius $r_B$, before rotation is:

$$\varphi'_{1;B} = m\psi + \beta m \qquad (5a)$$

And after rotation:

$$\varphi_{1;B} = m\psi + \beta m + \alpha \qquad (5b)$$

The phase of the second vortex, along the radius $r_B$:

$$\varphi_{2;B} = n\psi + \beta n \qquad (5c)$$

Along the $r_B$ line, the difference between these two-phase values is $\pi$, so after using (5):

$$\varphi_{1;B} - \varphi_{2;B} = m\psi + \beta m + \alpha - n\psi - \beta n = \pi \Rightarrow \beta = \frac{\alpha}{n-m} \qquad (6)$$

Figure 10 shows the case when $m = 3$ and $n = -2$.

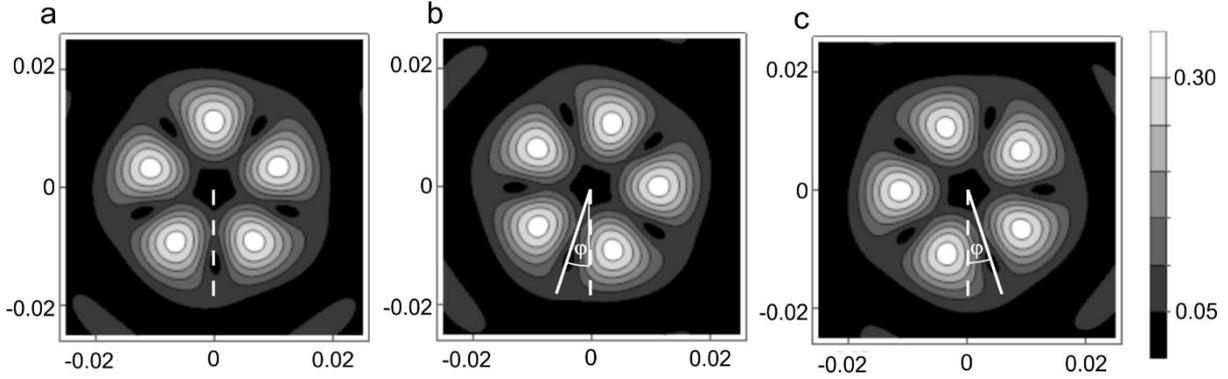

**Figure 10. Rotation of the higher-order vortex pattern. The topological charges are $m = 3$ and $n = -2$ a) the basic pattern, no additional phase is added; b) the phase equal $\alpha = \pi/2$ is added; c) the phase equal $\alpha = -\pi/2$ is added. As expected, the image has five-fold symmetry and the measured angle of rotation is equal to the added angle divided by 5, in this case $\varphi = \pi/10$.**

Figure 10 shows that change of the angle $\alpha$ rotates the whole pattern, with respect to the sing of $\alpha$. This is true even when the phase shift $\alpha$ is introduced to the second vortex pattern, which agrees with the formula (6). Figure 11 illustrates the case when both vortices have the topological charge of the same sign but different values. Redrawing Figure 8 for this example, one can notice that $|m| - |n|$ ($m > n$) symmetry lines are obtained. The formula (6) describing the rotation still holds (compare Figure 11d and f).

The light distribution between various parts of the image can be changed by modification of the pixel ratio from 1:1 to less symmetrical, which is illustrated in Figure 11. In this way, the week parts of the image can be made brighter. The reason for this is quite simple. Destructive interference between two vortex patterns occurs in the area of low intensity, whilst constructive interference is visible in the area of high intensity. Suppression of both interference effects is possible by reducing one of the components. Consequently, the bright parts may become weaker and dark parts stronger.

The well calibrated SLM gives a clear relation between the applied voltage to the SLM cell and the phase change of the light passing through. So, we do know what voltage should be applied to modulate the phase in the range of $2\pi$ for the given wavelength. If the error occurs the resulting phase is under or over modulated, which lowers the quality of the phase pattern reproduced on the SLM. It seems reasonable to assume that the formula (6) for angle rotation of the entire image was derived for perfectly calibrated SLM. However, this is not the case. As was shown in papers [42,43] the under or over-modulated SLM works in the same way as perfectly calibrated one, but with lower diffraction efficiency. We have also obtained the same rotation angles both numerically and experimentally, which means that the rotation angle is the same in case of broadband illumination (as for example under illumination from supercontinuum light source [44].

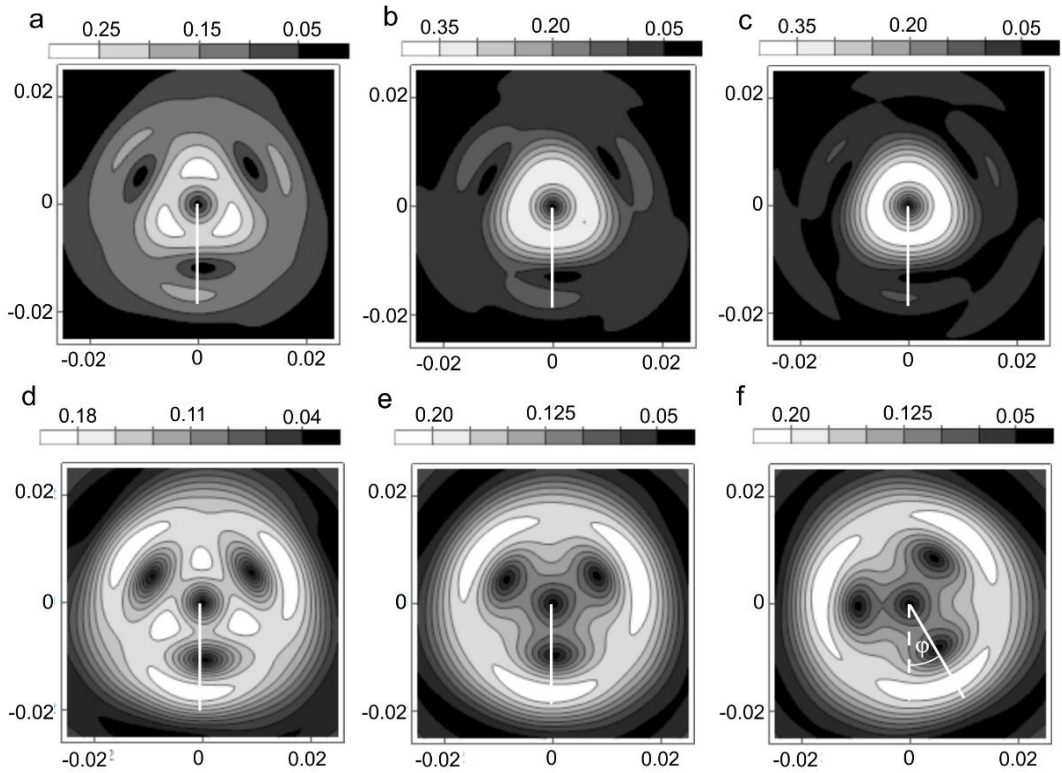

**Figure 11.** The interference images of two vortices having the topological charge of the same sign: $m = 4$ and $n = 1$. a) pixel ratio 1:1; b) pixel ratio 1:2 – the central part of the pattern is brighter. This is because the $n = 1$ vortex is built by twice a cells used for $m = 4$ vortex and the $n = 1$ vortex ring is smaller than $m = 4$, so the central part of the image is stronger; c) pixel ratio 1:3, the central part is even stronger; d) pixel ratio 2:1. Now the $m = 4$ vortex is built by twice cells used for $n = 1$ vortex. In this case, the outer part of the image is stronger; e) pixel ratio 3:1, the outer part is even more bright compared to the central part; f) pixel ratio 3:1 and the $n = 1$ pattern was rotated by $\pi/2$. The measured angle of the whole pattern rotation is $\varphi = \pi/6$ as calculated from the formula (6).

### 3. EXPERIMENT

The experimental setup is shown in

Figure 12. The SLM cells were divided randomly into two groups to display two various phase patterns representing two optical vortices. The optical quality of the SLM is relatively low. Hence, the phase correction map was implemented to improve it. This

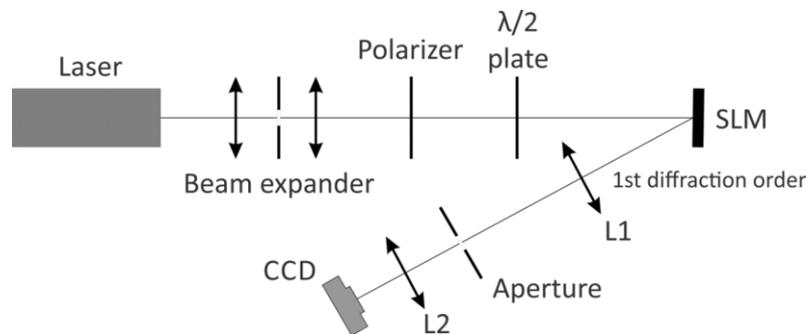

map was determined using combined interferometric-iterative methods described in [15].

**Figure 12. Experimental setup.** Expanded He-Ne laser beam ($< 5mW$, $\lambda = 632.8\,[nm]$), illuminates the SLM (Holoeye Pluto VIS-020). Polarizer together with half waveplate controls the polarization of incoming beam for proper SLM modulation. Due to low SLM diffraction efficiency the diffraction grating with period $350\,lines/mm$ was introduced. Thus, the vortex beam is embedded in the first order of diffraction and filtered by an aperture and telescope system: $L1 = 150\,[mm]$, $L2 = 75\,[mm]$. The final image is registered by CCD camera.

Figure 13 shows the experimental images. Figure 13a-c correspond to numerical calculations presented in Figure 4. The same rotation of the whole image, under the same phase shift of one of the vortex patterns, can be observed. Figure 14d-e show the image of the combined case $m = +4$, $n = -5$, and $m = +4$, $n = -9$, respectively. The image revels 9-fold and 13-fold symmetry as was expected from developed theory. Due to limited SLM resolution, the quality of Figure 13e is low. With increase of vortex topological charge, the produced phase map is more detailed.

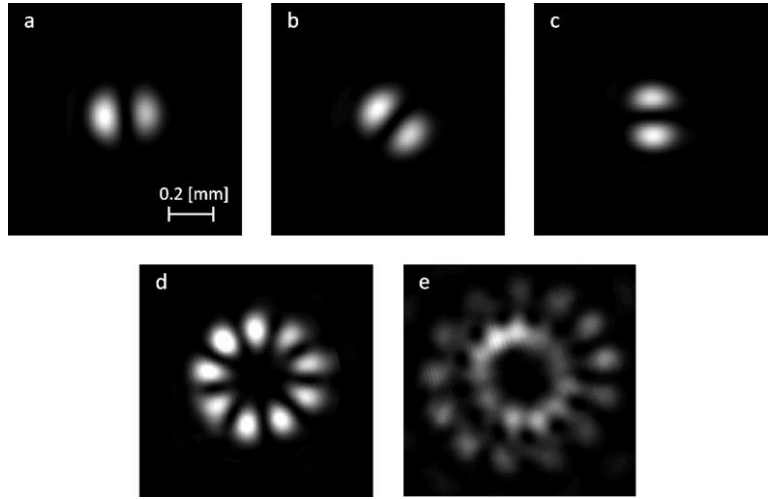

**Figure 13.** Examples of composite vortex patterns; a, b and c correspond to Figure 4 respectively. in the a) the vortex $m = +1$ and $n = -1$ are combined. In b) and c) the angle $\alpha = \pi/2$ and $\alpha = \pi$ is added respectively. Consequently, the image is rotated by $\alpha/2$. In d) the combination $m = +4$ and $n = -5$ is shown. The image reveals 9-fold symmetry. In e) the combination $m = +4$ and $n = -9$ is presented. Although the image is of poor quality, the 13-fold symmetry is visible.

Figure 15a-c show the rotation of the image produced by the m=+2 and n=-3 patterns. The total angle of rotation is in very good agreement with its calculated value. Figure 15d-f show the influence of the cells ratio. As was expected, some parts of the image becomes brighter or darker depending on the specific cell ratio The character of images changes are the same as in the case of numerical simulations. However, there are some small differences between theoretical and experimental results. Since generated structures are perfectly aligned with each other, the optical elements may be not. Which, together with the influence of the SLM quality can cause some of these differences.

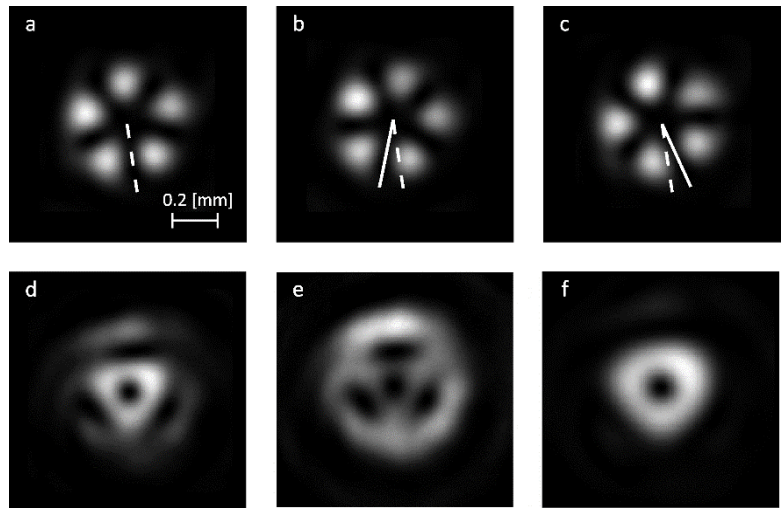

**Figure 14.** The gallery of combined vortices images. a-c shows the case of $m = +2, n = -3$. The extra phase angle $\alpha$ was added to the first vortex pattern ($m = +2$) as following: a) $\alpha = 0$, b) $\alpha = \pi/2$ and c) $\alpha = -\pi/2$. The white dashed line is a reference line indicating the zero line in the image. This reference line is copied to b) and c) showing the rotation of the entire pattern by angle $\pm \pi/10$, which can be calculated from the formula (6). d) Shows the case of $m = +1, n = +4$ with equal pixels distribution between both patterns. e) The same pattern, but the pixel ratio is 1:3, the external parts of the image d) became brighter and internal darker; f) the same pattern, but the pixel ratio is 3:1, the external parts of the image d) are practically invisible, whilst internal are brighter.

4. **CONCLUSIONS**

The composite vortex beams have attracted much attention during last twenty years. There are practical and purely scientific reasons for that. The composite beams serve for example to study the orbital angular momentum effects. They are used in optical trapping, optical fiber communications, and optical measurements. In this paper we have presented a stable and practical method of generating the whole family of composite vortex beams. Our method gives perfectly aligned beams, proofed against mechanical vibrations. If there are mechanical vibrations the both beams will jump in the same way and the resulting interference pattern will also jump at the screen but as a rigid body. This is not true in the case of interferometric methods. The method is easily available since the SLMs become more and more popular. Nowadays, the SLM quality and resolution, as well as several correction techniques, make them a reasonable choice for optical experiments. Similar superpositions could be achieved by a combination of Q-plates, waveplates, and polarizers. However, this approach will lack the possibility to control the relation between superposed structures and the alignment of these elements would largely influence the final image quality.

As we have shown, the composite vortex beams can be easily controlled. We can rotate them and change the distribution of light between various parts of the image. The calculated angles of rotation are independent of errors in the SLM calibration, so the theory can be also used for non-coherent illumination. The method has an obvious limit. It cannot be used if one of the vortex beams has to be modified separately, as for example, it is possible in Mach-Zehnder interferometric system.

The number of patterns which can be generated in this way is much wider than we have shown in this paper. We can mix pure Gaussian beam with a vortex beam as it was analyzed in [27], the Gaussian vortex beams can be replaced by Bessel beams or a vortex beams with any other amplitude envelope. We can shift a bit one of the beams against the other one by adding a grating pattern to one of the vortex patterns. We can slightly focus one of the beams by adding Fresnel lens phase pattern. We can mix three or more patterns, depending on the SLM resolution. We did not explore all these possibilities. Our goal was to present this technique on simple but important example. The question of its further enhancements is a matter of specific goals.

**FUNDING**

Polish Ministry of Science and Higher Education ("Diamond Grant") (DIA 2016 0079 45)
Nacional Science Centre (Poland) (UMO-2018/28/T/ST2/00125)